\newcommand{\ket}[1]{| #1 \rangle}
\newcommand{\bra}[1]{\langle #1 |}
\newcommand{\vxv}[1]{\ket{#1}\bra{#1}}
\newcommand{\brkt}[2]{\langle #1 \mid #2 \rangle}
\newcommand{\sqbk}[2]{\bigl| \langle #1 \mid #2 \rangle \bigr|^2}

\newenvironment{keywords}[1]{\begin{flushright}{\bf Keywords:} #1
                              \end{flushright}}

\documentclass{article}

\title{Analogue Quantum Computers for Data Analysis}

\author{Alexander Yu. Vlasov\footnote{E-mails:
  alex@protection.spb.su, vlasov@physics.harvard.edu}\\
Federal Center for Radiology, IRH \\
\small 197101, Mira Street 8, St.$\!$--Petersburg, Russia}
\date{29 December 1997 \\
11 February 1998}
\begin{document}
\maketitle
{\hfill \sf quant-ph/9802028}
\begin{abstract}
 Analogue computers use continuous properties of physical system for modeling.
In the paper is described possibility of modeling by analogue quantum computers
for some model of data analysis. It is analogue associative memory and a formal
neural network. A particularity of the models is combination of continuous
internal processes with discrete set of output states. The modeling of the
system by classical analogue computers was offered long times ago, but now
it is not very effectively in comparison with modern digital computers. The
application of quantum analogue modelling looks quite possible for modern
level of technology and it may be more effective than digital one, because
number of element may be about Avogadro number, $N \sim 6{\cdot}10^{23}$.
\end{abstract}



\section{Introduction}\label{chap:intr}

In the work \cite{vlasov:qcm96} was considered possible application of
analogue quantum computing to recognition of images. In present work the similar
approach is developed for some particular models. For the modeling are used
quantum systems with finite number of states and it is similar with
current models of ``digital'' quantum computers
\cite{feynman:comp,deutsch:turing,deutsch:gates}. Anyway the computations
are called here ``analogue'' because it is extension of models
supposed for realization by classical analogue computers already
20--30 years ago by different authors \cite{arbib:bmm,kohonen:assocmem}.

Feynman \cite{feynman:comp} has mentioned that using of quantum systems for
digital computation does not exploit all their properties.
Deutsch \cite{deutsch:turing,deutsch:gates} has used the features for
{\it quantum parallelism}, but they can be used also for some kind
of analogue computations.

\section{Quantum registers}\label{chap:qreg}

Let us consider a standard model of quantum computational network
\cite{deutsch:gates}. Some problems related with the approach is discussed
in \cite{vlasov:rqc}. For description of quantum system is used {\it finite
dimensional Hilbert space} ${\cal H}$, {\it i.e.} complex vector space with
Hermitian scalar product:
\begin{equation}
 \left({\bf a} , {\bf b} \right) = \sum_{i=1}^n a_i \overline{b_i} \ , \quad
 \| {\bf a} \| = \sqrt{\left( {\bf a}, {\bf a} \right)}
\label{eq:hermnorm}
\end{equation}
Due to complex conjugation of second vector this form sometime is called
sesquilinear. It would be bilinear, symmetrical scalar product {\it only}
for vectors with {\it all real} components.

There are physical notations \cite{dirac:pqm,feynman:flph3} $\ket{a}$ for
the vector  ${\bf a}$ and $\bra{a}$ for conjugated co-vector ${\bf a}^*$ ,
and
\begin{equation}
 {\bf b}^* {\bf a} = ({\bf a},{\bf b}) \equiv \brkt{b}{a} .
\label{eq:quantnorm}
\end{equation}

The vectors $\ket{\psi}$ and $\lambda \ket{\psi}$ for any nonzero complex
$\lambda$ are describe the same physical state. It means that states of
quantum systems are {\it rays} in the Hilbert space ({\it i.e.} elements of
{\it complex projective space} ${\bf C}P^n$ ). Due to the property we can
consider only states with unit norm $\|\psi\| = 1$.

A {\it qubit} \cite{schumacher:qubit,bennett:review} is a simple quantum
system with two states, it is described by 2D complex vector space. We can
choose two orthogonal vectors $\ket{0},\ \ket{1}$ as a basis of the space:
\begin{equation}
 \ket{0} \equiv \left( \matrix{1\cr 0} \right), \quad
 \ket{1} \equiv \left( \matrix{0\cr 1} \right)
\label{eq:qubit}
\end{equation}

A {\it quantum register} is a quantum system with N states. ``Binary'' quantum
computer uses some particular case of such {\it q-register}, {\it n-qubit
register} with $N = 2^n$, considered as compound quantum system with n qubits.
For analogue quantum computation under consideration it is possible, but not
necessary to use only the binary n-qubits, N could be any natural number.

A special property of quantum system is the process of {\it measurements}
\cite{feynman:flph3}.
It is impossible to get full information about quantum system. The simple
measurement device can be described by operator of {\it projection}
\begin{equation}
P_{\psi} \equiv \vxv{\psi}; \quad
P_{\psi} : \ket{\varphi} \to \ket{\psi}\brkt{\psi}{\varphi}
\label{eq:proj}
\end{equation}

The $P_{\psi}$ describes registration of a system $\ket{\varphi}$ with probability
$\sqbk{\psi}{\varphi}$. Here $\|\psi\| = \|\varphi\| = 1$, otherwise:
\begin{equation}
{\bf Prob}_{(\varphi \to \psi)} =
\frac{\sqbk{\psi}{\varphi}}{\brkt{\psi}{\psi}\brkt{\varphi}{\varphi}}
\label{eq:prob}
\end{equation}
The probability of registration is one only if
the $\ket{\psi}$ and $\ket{\varphi}$ describe the same physical state.
The probability is zero if $\ket{\varphi}$, $\ket{\psi}$ are orthogonal
vectors. The device has only two possibilities.

In more general case there are N {\it orthogonal} output states
$\ket{\psi_i}$. Let us have a quantum system $\ket{\chi}$
\begin{equation}
 \|\chi\| = 1, \quad \brkt{\psi_i}{\psi_j} = \delta_{ij}, \quad
 \ket{\chi} = {\sum_{i=1}^N a_i \ket{\psi_i}}, \; a_i = \brkt{\chi}{\psi_i} .
\label{eq:qubasis}
\end{equation}
We have one of $\ket{\psi_i}$ due to measurement of the system
$\ket{\chi}$ with probability $\Pr_i = | a_i |^2$. Here are
{\it N} versions of outcome instead of two in the previous example.

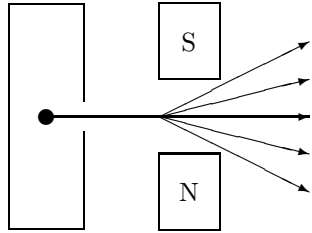
\begin{figure}[htbp]
\begin{center}
\unitlength=1.00mm
\linethickness{0.4pt}
\begin{picture}(40.00,35.00)
\put(5.00,20.00){\circle*{2.00}}
\put(10.00,18.00){\line(0,-1){13.00}}
\put(10.00,5.00){\line(-1,0){10.00}}
\put(0.00,5.00){\line(0,1){30.00}}
\put(0.00,35.00){\line(1,0){10.00}}
\put(10.00,35.00){\line(0,-1){13.00}}
\put(20.00,20.00){\vector(4,1){20.00}}
\put(20.00,20.00){\vector(4,-1){20.00}}
\put(20.00,20.00){\vector(2,-1){20.00}}
\put(20.00,20.00){\vector(2,1){20.00}}
\put(20.00,25.00){\framebox(8.00,10.00)[cc]{S}}
\put(20.00,5.00){\framebox(8.00,10.00)[cc]{N}}
\put(5.00,20.00){\vector(1,0){35.00}}
\end{picture}
\end{center}
\caption[Stern--Gerlach]{Stern--Gerlach experiment with a beam of particles}
\label{fig:stern}
\end{figure}

The simple example of the measurements is Stern--Gerlach device
\cite{feynman:flph3}, Fig.(\ref{fig:stern})\@.

\section{Data analysis}\label{chap:dataan}

For application to data analysis let us consider more simple case of
real vector space ${\bf R}^n$ (or projective space ${\bf R}P^n$)
\cite{vlasov:qcm96}. The scalar product is:
\begin{equation}
 \left( \vec{w},\vec{v} \right) \equiv  \sum_{i=1}^n w_i v_i
 \label{eq:norm}
\end{equation}
The norm Eq.(\ref{eq:norm}) is real version of complex norm used in
Eqs.(\ref{eq:hermnorm},\ref{eq:quantnorm},\ref{eq:proj}). For vectors
with unit length it is just a cosine of the angle between them. The
bilinear measure is often used for data analysis. An application to
image recognition was described in \cite{vlasov:qcm96}.

Historically, one of possible reasons for considering such kind of models was
observation of properties of usual biological neuron \cite{arbib:bmm}. The
formal model of neuron \cite{sokolov:neuron} can be used for convenient
description of analogue data analysis under consideration. The formal neuron
sums input signals of synapses $s_i$ with some weights $w_i$, see Eq.(\ref{eq:norm}),
$A = \left( \vec{w},\vec{s} \right) = \sum_i w_i s_i$. Here $\vec{w}$ is a
vector of weights and $\vec{s}$ is a vector of signals.

In models of artificial neural networks can be used some additional
transformations $f : A \to f(A)$. In one of the first
model, Rosenblatt's {\it perseptron}, it was threshold function $f(A) =
\theta(A - A_0)$, {\it i.e.} $f(A)$ is 1 if $A > A_0$ and otherwise it is zero.
An application of possible realization of similar system by quantum computer
for expert systems was described shortly in \cite{vlasov:expsys}.

In many contemporary work $f(A)$ is {\it nonlinear}\footnote{A nonlinearity is
bad for realization by analogue quantum computers} functions \cite{kosko:fuzzy},
like $s(A) = 1/(1+e^{-c A})$. The using of nonlinear model was related with
a problem of modeling arbitrary binary functions by perseptron
\cite{minsky:nogo}. But it is not the problem for quantum
networks \cite{deutsch:gates,feynman:comp}.

Let us use algorithms without using of nonlinear functions,
\cite{sokolov:neuron,kohonen:assocmem}. The models are useful
as an introduction to analogue quantum computations discussed further.

\medskip

In the \cite{sokolov:neuron} is considered method of {\it coding a value
of parameter by number of channel} (CVPNC), i.e. number of (most) activated
element of network corresponds to value of parameter. The work uses network
of the formal neurons with equal length of weight vectors $|\vec{w}(k)|$.
Maximum of output signal $ A(k) = \left(\vec{w}(k),\vec{s}(k)\right)$ corresponds
to the same direction of the vectors $\vec{s}$ and $\vec{w}$. Such kind of
network does not distinguish the vectors $\vec{s}$ and $\alpha\vec{s}$ for
a positive real $\alpha$. The space of signals is similar with space of ray
discussed in section {\bf\ref{chap:qreg}}.

\medskip

A model of analogue auto-associative memory (AAAM) is presented in
\cite{kohonen:assocmem}. For representation of images is used vector
space ${\bf R}^N$. In chapter 2.3 of the book is described error correction
for images by orthogonal projection to subspace, linear span of the images.
As an example is used set of gray scaled pictures with $N = n \times m$ elements.
The best result is produced for orthogonal images. If number of images
$i \ll N$ then they are ``almost orthogonal'' because scalar product of
two random vectors with unit norm is:
\begin{equation}
 \left(\vec{w},\vec{v} \right) \sim N^{-1/2}
 \stackrel{N \to \infty}{\longrightarrow} 0
\label{eq:scalz}
\end{equation}

\section{Quantum analogue computations}\label{chap:quantan}

Two last examples in previous section can be considered as a real basis
for analogue quantum modeling because the basic operations are similar with
measurement processes described by Eqs.(\ref{eq:proj},\ref{eq:qubasis})
and discussed above in relation with the formulae.

In case of quantum register we have a complex space instead of real, but
it is possible to use the models. First model, {\sf CVPNC}, uses property
of maximum scalar product Eq.(\ref{eq:norm}) for two equal vectors on hypershpere.
The Hermitian norm Eq.(\ref{eq:hermnorm}) has the same property and probability
of registration ${\bf Prob}_{(\varphi \to \psi)}$ in Eq(\ref{eq:prob}) is
maximum, {\it unit}, if $\ket{\varphi}$ and $\ket{\psi}$ are the same physical
state (the same {\it ray} in Hilbert space).

\medskip

It is possible to describe the quantum {\sf CVPNC} model : We have set of
properties. Each property described by a vector (ray). An analogue system
must find number of property corresponded to input signal. It can be
described also as ``{\it recognition} of an image stored in associative
memory of the system'' (see \cite{vlasov:qcm96}).

If all vectors are orthogonal, it is enough to use one device described by
Eq.(\ref{eq:qubasis}) with $\ket{\psi_i}$ is equal to $i$-th vector in the
set of properties (images). If input signal corresponds to one of stored
properties, the output channel will have desired number with certainty, but
if the signal have some error the process would be probabilistic. For small
errors the most probability has desired channel.

There is a problem if signal differ enough with all images. For example,
if the vector of signal is composition of two images, $i$-th and $j$-th,
$\ket{s} = \alpha \ket{w(i)} + \beta \ket{w(j)}$, ($|\alpha|^2+|\beta|^2 = 1$
because norm of $\ket{s}$ is unit) then the output would be either $i$
with probability $|\alpha|^2$ or $j$ with probability $|\beta|^2=1-|\alpha|^2$.

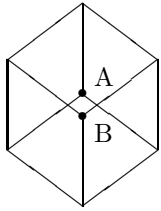
\begin{figure}[htbp]
\begin{center}
\begin{minipage}{3cm}
\unitlength=0.50mm
\linethickness{0.4pt}
\begin{picture}(40.00,59.00)
\put(0.00,44.00){\line(4,3){20.00}}
\put(20.00,59.00){\line(4,-3){20.00}}
\put(40.00,44.00){\line(-4,-3){20.00}}
\put(20.00,29.00){\line(-4,3){20.00}}
\put(0.00,20.00){\line(4,3){20.00}}
\put(20.00,35.00){\line(4,-3){20.00}}
\put(40.00,20.00){\line(-4,-3){20.00}}
\put(20.00,5.00){\line(-4,3){20.00}}
\put(20.00,29.00){\circle*{2.00}}
\put(20.00,35.00){\circle*{2.00}}
\put(23.00,37.00){\makebox(0,0)[lb]{A}}
\put(23.00,27.00){\makebox(0,0)[lt]{B}}
\put(0.00,44.00){\line(0,-1){24.00}}
\put(40.00,44.00){\line(0,-1){24.00}}
\put(20.00,29.00){\line(0,-1){24.00}}
\put(20.00,59.00){\line(0,-1){24.00}}
\end{picture}
\end{minipage} \ %
\begin{minipage}{8cm}
{\small
The picture of a cube is an example of ambiguous image recognition. It
is impossible to find with certainty that point is nearer, A or B. Sometimes
the picture looks like cube with upper face is close to observer, sometimes
with lower one. Sometimes it can looks like a flat figure. }
\end{minipage}
\end{center}
\caption[Ambiguous picture]{Ambiguous picture of a cube}
\label{fig:cube}
\end{figure}

But this is not specific problem of the system under consideration, the
human vision has the similar kind of behavior for an ambiguous images,
see Fig.(\ref{fig:cube}). For modeling such system by usual computer it
would be necessary to use random number generator, but for the system under
consideration such results was made ``free'', without any extra modules.

\medskip

One of simple examples of such a system could be Stern--Gerlach device,
Fig.(\ref{fig:stern}). Here an input signal is coded by state of particle
in the beam \cite{vlasov:qcm96}. A disadvantage of the device is macroscopic
size. There is some problem with description of microscopic version of processes
related with irreversible operators of projection $P_{\psi}$. On the other
hand such processes are quite common and they are considered as one of main
problems for quantum computers (reduction, decoherence, decay, etc.). Contrary,
the processes could be useful for analogue quantum computations described
above.

\medskip

If vectors of images are not orthogonal it is impossible to use the simple
model with one quantum system. An extra problem is because evolution of quantum
system described by unitary operators. The operators {\it by definition}
can not change the scalar product Eqs.(\ref{eq:hermnorm},\ref{eq:quantnorm}).
So if two vectors are not orthogonal, it is impossible to correct the problem
by unitary quantum evolution.

An advantage of orthogonality was discussed in \cite{kohonen:assocmem}
for {\it real} vector space and {\sf AAAM}. There is a property of error
correction by orthogonal projection to linear span of all images. For
analogue quantum computation it is possible to use similar way of error
correction by projection on subspace. It is similar to error
correction schemes of ``digital'' quantum computers\cite{peres:corr}.

For complex vector spaces it is also possible to use property similar to
Eq.(\ref{eq:scalz}), {\it i.e.} ``almost orthogonality'' of random rays.

The using of nonorthogonal images is possible if input signal is represented
by few quantum systems in the same state instead of only one system. An
example is Stern--Gerlach experiment with filtered beam of atoms as an input%
\footnote{Various kinds of such experiments are described in \cite{feynman:flph3}}.
It may be possible to distribute the signal between few {\it filters}
$P_{w(k)}$, see Eq.(\ref{eq:proj}), for realization quantum version of
{\sf CVPNC} model (see also \cite{vlasov:qcm96}).

\section{Conclusion}

In the paper is presented some possible applications of analogue quantum
computations. For the modeling was used quantum system with finite number
of states. It is shown that such kind of system could be used for development
of known models like associative memory by T.Kohonen (computer science) and a
model of neural network by E.Sokolov (experimental neurophysiology).

In a simplest case it is enough to use one system like atom with a few
states for each element. It make possible to create media with number of
elements about $N \sim 10^{23}$. One of disadvantages of the model is
using a data from associative memory instead of more advanced algorithmic
data analysis. But it can be compensated for some application by such huge
amount of elements. It is similar to using of reflexes \cite{sokolov:neuron}
in standard situations instead of reasoning.

\section*{Acknowledgments}
I am grateful to Prof. Andrew Grib, Roman Zapatrin, Sergey Krasnikov,
and other participants of seminars in Friedman Laboratory for Theoretical
physics for discussions about quantum computers. The help of Andrew Borodinov
with artificial neural networks, parallel distributive processing, fuzzy
logic, associative memory, {\it etc.} was very useful for me.


\end{document}